\documentclass[a4paper]{jpconf}
\usepackage{graphicx}
\usepackage{amsbsy}
\usepackage{color}
\usepackage{amsmath}
\usepackage{amssymb}
\begin{document}
\title{Relic neutrino clustering in the Milky Way}

\author{Pablo F. de Salas}

\address{The Oskar Klein Centre for Cosmoparticle Physics, Department of Physics,
Stockholm University, AlbaNova, 10691 Stockholm, Sweden}

\ead{pablo.fernandez@fysik.su.se}

\begin{abstract}
The Standard Cosmological Model predicts the existence of relic neutrinos, which are indirectly probed through the effective number of relativistic species in the early Universe. In addition, from neutrino flavour oscillations we know that at least two of the neutrino mass states have a non-zero mass. Since the expansion of the Universe has diluted the energy of relic neutrinos, those that are massive are also non relativistic today. This means that they can be trapped in strong gravitational potentials, such as the one of the Milky Way. We review the calculation of the local overdensity of relic neutrinos produced from the gravitational attraction of our Galaxy, Andromeda and the Virgo cluster, commenting on the implications for an experiment aiming at relic neutrino detection, such as the PTOLEMY project.
\end{abstract}

\section{Introduction}
Current precision of most neutrino oscillation parameters is at the few per cent level. In particular, neutrino oscillations tell us that there are at least two neutrino states with a non-zero mass of at least $\sim 9\,\mathrm{meV}$ \cite{deSalas:2017kay}.
On the other hand, the standard model of cosmology, the so-called $\Lambda$CDM model, predicts that the current kinetic energy of relic neutrinos, which decoupled from the cosmic bath when the Universe was barely $\sim 1\,\mathrm{s}$ old, has diluted because of the cosmic expansion and now average $\sim 0.17\,\mathrm{meV}$, much smaller than the minimum required mass of the massive neutrino states.
For this reason we know that at least two relic neutrino species are non relativistic nowadays.
Therefore, they can cluster under the influence of strong gravitational potentials, such as the one of our Galaxy.

Although relic neutrinos are by far the most common neutrinos in the Universe (its mean number density is 
$n_{\nu,0} \sim 56\,\mathrm{cm^{-3}}$ per neutrino flavour and chirality), 
not a single relic neutrino has been detected directly. 
We know of their existence in an indirect way from 
$N_{\rm eff}$,
a cosmological parameter that accounts for all radiation at the early Universe that does not come from photons, 
and whose observed value \cite{Aghanim:2018eyx} is compatible with the standard theoretical estimate 
$N_{\rm eff} = 3.045$ \cite{Mangano:2005cc,deSalas:2016ztq,Gariazzo:2019gyi}. 
However, since neutrinos only participate in weak interactions, and relic neutrinos in particular have such a low kinetic energy, these relics are extremely difficult to catch.

At present, the best suggested method to detect relic neutrinos directly is to look for the signature caused by their capture in atomic tritium \cite{Weinberg:1962zza,Cocco:2007za}, a peak that will be shifted about two times the mass of the neutrino from the true $\beta$-decay endpoint.
This is the strategy chosen by the PTOLEMY collaboration \cite{Betts:2013uya,Baracchini:2018wwj}, which has recently presented the neutrino physics program that will be studied by the project \cite{Betti:2019ouf}.
Since the relic neutrino capture event rate depends on the local number density of relic neutrinos, $n_{\nu, \odot}$, 
a key aspect in order to interpret adequately a future direct detection of these particles relies on 
knowing the exact overdensity that is due to gravitational attraction at our location in the Milky Way.

\section{Computing the relic neutrino clustering}
We are going to cover two different ways of computing the local density of relic neutrinos,
following the works of \cite{deSalas:2017wtt} and \cite{Mertsch:2019qjv}.
Both studies are based on the $N$-one-body simulations introduced in \cite{Ringwald:2004np}.
While the first work \cite{deSalas:2017wtt} follows a traditional forward-tracking of the trajectories of the $N$ simulated test particles,
the second work \cite{Mertsch:2019qjv} exploits the back-tracking technique that is common in cosmic ray propagation.

\subsection{$N$-one-body simulations}
The idea behind $N$-one-body simulations \cite{Ringwald:2004np,deSalas:2017wtt,Zhang:2017ljh,Mertsch:2019qjv} is that we can compute the overdensity of relic neutrinos by following the evolution of many ($N$) independent (one-body) particles.
This simplified approach is possible provided that \emph{a)} the evolution of each of the test particles that we compute is not affected by the evolution of the other particles of the same type, and \emph{b)} the evolved test particles have no impact on the evolution of the matter potential of the system.
Since our test particles are relic neutrinos (whose masses are extremely small to have a gravitational influence and they interact very weakly), and our system is the Milky Way (plus eventually Andromeda and the Virgo cluster), the previous requirements are effectively satisfied.
Therefore we can follow the trajectory of $N$ individual test particles in the same evolving gravitational potential, sampling their initial or final phase space depending on whether the simulation is based on a forward- or a back-tracking technique, respectively. Finally, we compute the local density of relic neutrinos by associating a corresponding weight to the initial phase-space of the relevant test particles in the forward-tracking approach, 
or by integrating the final distribution function over the neutrino momentum in the back-tracking approach.





\subsection{Forward- versus back-tracking}

Both forward- and back-tracking techniques have their own advantages and disadvantages, but in both cases we assume the initial phase space of relic neutrinos (at an initial redshift of $z\geq 3$) to follow a homogeneous and isotropic Fermi-Dirac distribution. 
This distribution is then perturbed at earlier times, when the simulated test particles are trapped in the growing gravitational potential of the relevant structures that we consider, either the Milky Way alone or in combination with Andromeda and the Virgo cluster.

The main difference between the two approaches is the direction of the time arrow in our simulations.
In \cite{deSalas:2017wtt} (based on a forward-tracking approach), we start our simulations at the \emph{initial} redshift $z=3$. Then we follow the trajectories of $N$ individually simulated particles up to the final redshift $z=0$. 
On the contrary, in \cite{Mertsch:2019qjv} (based on a back-tracking approach) we invert the arrow of time, starting our simulations at the \emph{final} redshift $z=0$ and following the trajectories backwards in time.
Therefore, when we use the forward- or back-tracking technique we sample, respectively, the initial or final phase space of relic neutrinos. 
In the back-tracking approach we use Liouville's theorem to connect the final phase space with its homogeneous and isotropic initial counterpart.


Since we are interested in the local density of relic neutrinos (i.e. at the position of the Earth in the Milky Way), a back-tracking technique allows us to reduce the number of phase-space variables that we must sample over.
This advantage comes from the fact that, when back-tracking the trajectories of the test particles, we can fix the final position already at the location of the Earth in the Galaxy, sampling only over the three final momenta.
In addition, this allowed us in \cite{Mertsch:2019qjv} to relax the approximation of spherical symmetry that was needed in \cite{deSalas:2017wtt} in order to reduce the computational time.
Thanks to the relaxation of this symmetry we could include in our simulations the gravitational effect of Andromeda and the Virgo cluster.

\section{Modelling the gravitational potentials}

In both studies \cite{deSalas:2017wtt} and \cite{Mertsch:2019qjv} we model
the gravitational potential of the Milky Way as made of a dark matter (DM) halo and a baryonic contribution in form of a bulge and double exponential discs.
The shape of the baryonic profiles are inspired by the observational analysis \cite{Misiriotis:2006qq},
but notice that in \cite{deSalas:2017wtt} we further spherically symmetrised the contribution from baryons.
On the other hand, 
we model the DM halo with two different parameterisations in order to test the impact of uncertainties in this component: a Navarro-Frenk-White (NFW) and an Einasto profile.
We fit the DM parameters to the energy density $\rho_{\rm DM}(R)$ estimates at different distances $R$ from the Galactic centre obtained in \cite{Pato:2015tja}.
Recent observations from the ESA/Gaia mission \cite{Brown:2018dum} have improved enormously our knowledge of the Galaxy, and the Milky Way's rotation curve is determined with incredible precision (see e.g. \cite{Eilers:1810.09466} and an analysis of its determined circular velocity curve in order to estimate $\rho_{{\rm DM},\odot}$ in \cite{deSalas:2019pee}).
However, in order to set the value of the mass in the DM profiles we need to go beyond the radial extent of the analysis \cite{Pato:2015tja} and even that of \cite{deSalas:2019pee}.
Therefore, to set the DM mass we also use information of Milky Way satellites that can be as distant as $\sim 300\,\mathrm{kpc}$ from the Galactic centre \cite{Watkins:2010fe}.


For the Andromeda galaxy and the Virgo cluster, accounted for in \cite{Mertsch:2019qjv} together with the Milky Way, we only consider a spherical NFW profile, neglecting the baryonic content of Andromeda.

The redshift evolution of the DM components of the Milky Way, Andromeda and Virgo is taken into account following the $N$-body trend from \cite{Dutton:2014xda}, with a fixed virial mass and evolving virial concentration and radius.
The baryonic content of the Milky Way, however, is evolved in redshift assuming only a changing total mass, according to the averaged trend found in eight Milky Way-like galaxies from $N$-body simulations presented in \cite{Marinacci:2013mha}.

For further details on how we model the relevant gravitational potentials we refer the reader to the original publications \cite{deSalas:2017wtt,Mertsch:2019qjv}.


\section{Results and implications for a PTOLEMY-like experiment}

The local density of relic neutrinos $n_{\nu,\odot}$ varies depending on the mass of the neutrino $m_\nu$.
As an example, a neutrino with $m_\nu = 60\,\mathrm{meV}$ would have an enhancement of $\sim \text{10--20}\%$ from $n_{\nu, 0}$, while the enhancement for $ m_\nu = 150\,\mathrm{meV}$ can be up to $\sim 200\%$.
However, for a fixed neutrino mass, $n_{\nu,\odot}$ is found to be mostly dependent on the DM profile of the Milky Way. In particular, it is conditioned, as expected, by the Galactic mass contained within $R_\odot$.

\begin{figure}[t!]
\centering
\includegraphics[width=0.7\textwidth]{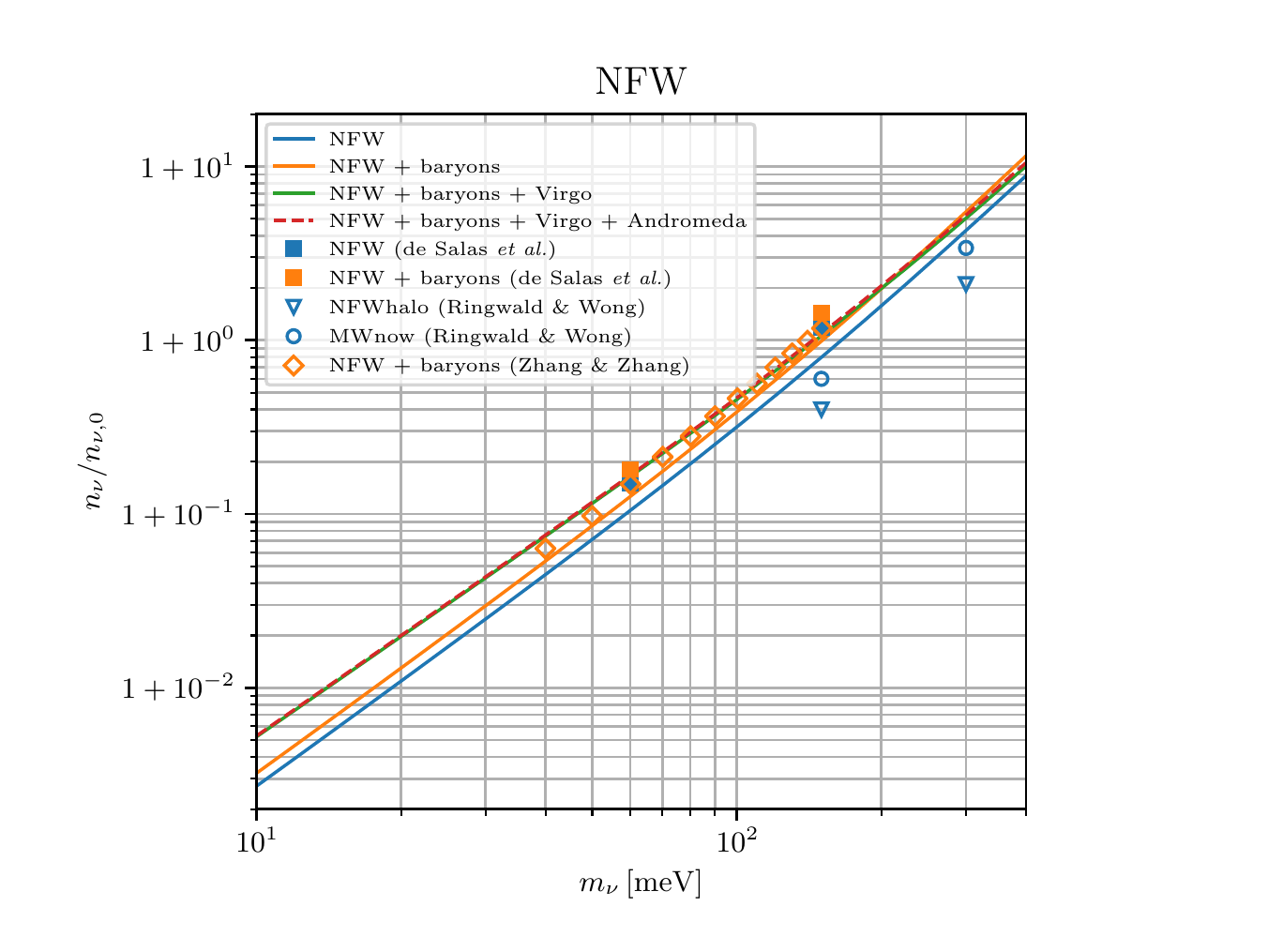}
\caption{Local neutrino overdensity depending on the neutrino mass $m_\nu$, when the DM halo of the Milky Way is assumed to have a NFW shape.
Lines correspond to different scenarios discussed in \cite{Mertsch:2019qjv}, including the contribution from baryons, Andromeda and Virgo.
The results from other references \cite{Ringwald:2004np,deSalas:2017wtt,Zhang:2017ljh} are also shown in a comparative basis.
Figure taken from \cite{Mertsch:2019qjv}.
}\label{fig:resumen-NFW}
\end{figure}

Including Andromeda does not affect much $n_{\nu, \odot}$, but the Virgo cluster has a non-negligible effect (see Fig. \ref{fig:resumen-NFW}) that is more pronounced the smaller the value of $m_\nu$. 
Interestingly, for larger $m_\nu$ the presence of Virgo \emph{reduces} the value of $n_{\nu, \odot}$ instead of enhancing it.
This behaviour occurs when $m_\nu \gtrsim 200\,\mathrm{meV}$ and is a result of the neutrino halo being compressed when neutrinos become colder (because they have a larger mass). Therefore, some neutrinos that would have ended up at our location in the Galaxy fall instead in the Virgo gravitational potential.

Knowing the exact value of $n_{\nu, \odot}$ is important to reduce degeneracies in the event rate $\Gamma_{\rm C\nu B}$ predicted in a PTOLEMY-like experiment \cite{Betti:2019ouf}. In particular, there is a factor of two in $\Gamma_{\rm C\nu B}$ that depends on the neutrino nature. If neutrinos are Majorana particles, twice as many of these particles are expected to be detected in a PTOLEMY-like experiment based on the capture of relic neutrinos in tritium \cite{Long:2014zva}.
Unfortunately, given current constraints to neutrino masses from cosmology \cite{Aghanim:2018eyx}, we expect a relic neutrino local overdensity of just a few tens of per cent over their vacuum number density.


\section*{Acknowledgments}
Work supported by the Vetenskapsr{\aa}det (Swedish Research Council) through contract No. 638-2013-8993 and the Oskar Klein Centre for Cosmoparticle Physics.

\section*{References}

\end{document}